\documentclass[twocolumn,showpacs,amsmath,amssymb,aps,pra,floatfix]{revtex4}
\usepackage{bm}
\usepackage{amsfonts}
\usepackage{graphicx}
\begin{document}
\title{Pinning and transport of cyclotron (Landau) orbits by electromagnetic vortices}
\author{Iwo Bialynicki-Birula}
\email{birula@cft.edu.pl}
\affiliation{Center for Theoretical Physics, Polish Academy of Sciences, Al. Lotnik\'ow 32/46, 02-668 Warsaw, Poland and\\
Institute of Theoretical Physics, Warsaw University, Ho\.za 69,
00-681 Warsaw, Poland}
\author{Tomasz Rado\.zycki}
\email{torado@fuw.edu.pl} \affiliation{Physics Department, Warsaw
University, Ho\.za 74, 00-682 Warsaw, Poland}

\begin{abstract}
Electromagnetic waves with phase defects in the form of vortex lines
combined with a constant magnetic field are shown to pin down
cyclotron orbits (Landau orbits in the quantum mechanical setting)
of charged particles at the location of the vortex. This effect
manifests itself in classical theory as a trapping of trajectories
and in quantum theory as a Gaussian shape of the localized wave
functions. Analytic solutions of the Lorentz equation in the
classical case and of the Schr\"odinger or Dirac equations in the
quantum case are exhibited that give precise criteria for the
localization of the orbits. There is a range of parameters where the
localization is destroyed by the parametric resonance. Pinning of
orbits allows for their controlled positioning. They can be
transported by the motion of the vortex lines.
\end{abstract}
 \maketitle

\section{Introduction}

The transverse motion of charged particles in a constant magnetic
field is fully delocalized. The classical trajectories of particles
projected on the plane perpendicular to the field are circles which
can be moved around without changing the particle energy. The wave
functions of a particle in quantum mechanics exhibit the same
behavior. If $\psi({\bm r})$ is a solution of the Schr\"odinger or
the Dirac equation with the energy $E$, then a shifted wave function
$\exp(ie{\bm r}\!\cdot\!({\bm r}_0\times{\bm B})/2\hbar)\psi({\bm
r}-{\bm r}_0)$ is also a solution with the same energy. In the
present paper we shall show that electromagnetic beams with vortex
lines localize the classical and quantum states at the position of
the vortex and when the vortex line is moved the orbits will follow.
In order to describe these properties quantitatively we shall use
new exact solutions of the equations of motion obtained in the
presence of an electromagnetic field comprising a constant magnetic
field and a wave with a vortex line. This paper extends our earlier
analysis \cite{prl} by adding a constant magnetic field. This
extension is significant because the addition of a constant field
introduces the third parameter characterizing the electromagnetic
field. From the three parameters (the wave frequency and two field
amplitudes) we can construct two dimensionless parameters and the
space of distinct solutions, as compared to those given in
\cite{prl}, becomes much richer. An extensive analysis of the exact
solutions of relativistic equations of motion can be found in a
monograph by Bagrov and Gitman \cite{bg}. However, they overlooked
the existence of the solutions described in this paper. A related
problem --- the motion of a particle in a constant magnetic field
and a plane electromagnetic wave --- has been treated in detail by
Roberts and Buchsbaum \cite{rb} a long time ago. The crucial
difference between these two cases is that the translational
symmetry in the plane perpendicular to the magnetic field is broken
by the presence of an electromagnetic vortex and the problem becomes
truly three-dimensional.

The aim of the present paper is not only to present new analytic
solutions of the classical and quantum equations of motion in some
realistic configurations of the electromagnetic field but also to
describe a universal confining mechanism of charged particles that
might have experimental applications. This mechanism acts for all
electromagnetic waves with a definite angular momentum. Such waves
are characterized by the presence of an electromagnetic vortex. At
the vortex the field vanishes and near the vortex the electric and
magnetic field vectors exhibit a characteristic behavior that leads
to the trapping of particles. The trapping of atoms by
electromagnetic vortices is a well established phenomenon
\cite{ahd,freegarde,slm}. It employs the dipole force pushing atoms
in the direction of a decreasing field amplitude in a laser beam
that is blue-detuned from the relevant atomic transition. The
mechanism for trapping charged particles by electromagnetic vortices
is completely different. It employs the rotation of the electric and
magnetic field vectors near the vortex line.

This paper contains a complete, classical, and quantum mechanical
description of the motion of charged particles in a combination of
an electromagnetic wave with an embedded vortex and a constant
magnetic field. Our treatment is fully relativistic but we also
present the results in the nonrelativistic limit. In Sec. II we
introduce a model electromagnetic wave that approximates very well,
near the vortex line, all solutions of Maxwell equations with the
lowest angular momentum. In Sec. III we give general solutions of
the Lorentz equations in such a wave and exhibit a mechanism that
confines the particle in the vicinity of the vortex line. In Sec. IV
we show how trapped particles can be transported by moving vortex
lines. In Sec. V we carry our analysis to the quantum theory
presenting exact solutions of the Dirac equation in the presence of
a wave with a vortex and the magnetic field. Finally, we present
solutions of the Schr\"odinger equation describing the transport of
Landau orbits.

\section{The simplest electromagnetic beam with a vortex line}

The notion of a vortex line of the electromagnetic field can be
traced back to an old paper by Rainich \cite{rainich}. A detailed
analysis with many examples has been recently given by us in Ref.
\cite{emvort}. In thia paper we shall use only a very special case
of the vortex line --- the one that is found in null solutions of
the Maxwell equations \cite{kiev}. In reality, vortex lines of the
electromagnetic field are most commonly found in beams carrying
angular momentum \cite{soskin,apb,mcgd}. Such beams can now be
produced with the use of several methods: computer generated
holograms, axicons, spatial light modulators, and biaxial crystals,
etc.. The best known examples of such beams are the Bessel beams and
the exact Laguerre-Gaussian (LG) beams. These are solutions of the
Maxwell equations obtained by separating the variables in
cylindrical coordinates \cite{stratton}. A convenient description of
these beams in terms of just one complex function is obtained with
the use of a complex vector
\begin{eqnarray}\label{rs}
{\bm F}({\bm r},t) = {\bm E}({\bm r},t)+ic{\bm B}({\bm r},t),
\end{eqnarray}
which we named the Riemann-Silberstein (RS) vector
\cite{pio,keller}. The RS vector obeys the complex form of Maxwell
equations
\begin{eqnarray}\label{max}
\partial_t{\bm F}({\bm r},t) = -ic\nabla\times{\bm F}({\bm r},t),\;\;\nabla\!\cdot\!{\bm F}({\bm r},t)=0.
\end{eqnarray}
We shall use a representation of ${\bf F}$ in terms of just one
complex function
\begin{subequations}\label{solf}
\begin{align}
F_x &= (\partial_x\partial_z + \frac{i}{c}\partial_y\partial_t)\chi,\\
F_y &= (\partial_y\partial_z - \frac{i}{c}\partial_x\partial_t)\chi,\\
F_z &= -(\partial_x^2+\partial_y^2)\chi.
\end{align}
\end{subequations}
where $\chi$ is an arbitrary complex solution of the wave equation
\begin{eqnarray}\label{wave}
(\frac{1}{c^2}\partial_t^2-\Delta)\chi({\bm r},t) = 0.
\end{eqnarray}
An equivalent representation of the electromagnetic field expressed
in terms of two real solutions of the wave equation, instead of a
single complex solution of the wave equation, has been given by
Whittaker \cite{whitt}. The functions $\chi({\bm r},t)$ for the
Bessel beams and for the Laguerre-Gaussian beams are most easily
expressed (cf. \cite{bb}) in the cylindrical coordinates
$(\rho,\phi,z)$. The Bessel beams are labeled by the following
parameters (``quantum numbers''): the transverse wave vector
$k_\perp=\sqrt{k_x^2+k_y^2}$, the (dimensionless) angular momentum
$m$, the wave vector $k_z$ in the direction of propagation, and the
helicity $\sigma=\pm 1$
\begin{align}\label{bes}
&\chi_B(\rho,\phi,z,t) = e^{-i\sigma(\omega_kt- k_z z -
m\phi)}J_{m}(k_\perp\rho).
\end{align}
The Laguerre-Gaussian beams are defined as
\begin{align}\label{lg}
&\chi_{LG}(\rho,\phi,z,t)\nonumber\\
& = \frac{e^{-i\sigma(\omega t_- -m\phi)}\rho^m}
{a(t_+)^{n+m+1}}\exp\left(-\frac{\rho^2}{a(t_+)}\right)
L_n^m\!\left(\frac{\rho^2}{a(t_+)}\right),
\end{align}
where $t_\pm=t\pm z/c$ and $a(t_+)=l^2+i\sigma c^2t_+/\omega$, and
$L_n^m$ is the Laguerre polynomial. The natural number $n$ gives the
number of zeros of the polynomial but otherwise has no direct
physical meaning. The parameter $l$ determines the transverse size
(waist) of the LG beam. In the limit, when $k_\perp\to 0$ for Bessel
beams and $l\to\infty$ for LG beams, both functions (\ref{bes}) and
(\ref{lg}) reduce (for $m>0$ and apart from numerical prefactors) to
the same solution of the wave equation
\begin{eqnarray}\label{same}
\chi({\bm r},t) = (x+iy)^m e^{-i\sigma\omega(t-z/c)}.
\end{eqnarray}
This choice of $\chi$ leads to the following RS vector
\begin{eqnarray}\label{rs1}
{\bm F}({\bm r},t) = \sigma \left({\hat{\bf x}}+i{\hat{\bf
y}}\right) (x+iy)^{m-1} e^{-i\sigma\omega(t-z/c)},
\end{eqnarray}
where we again disregarded some numerical prefactors. The RS vector
(\ref{rs1}) is a good approximation to the Laguerre-Gaussian and
Bessel beams near the $z$ axis, when $\rho\ll l$ or $\rho k_\perp\ll
1$. It belongs to a family of the solutions of the Maxwell equations
describing vortex lines riding atop a null electromagnetic field
\cite{kiev}.

\begin{figure}[ht]
\centering {\includegraphics[width=0.45\textwidth]{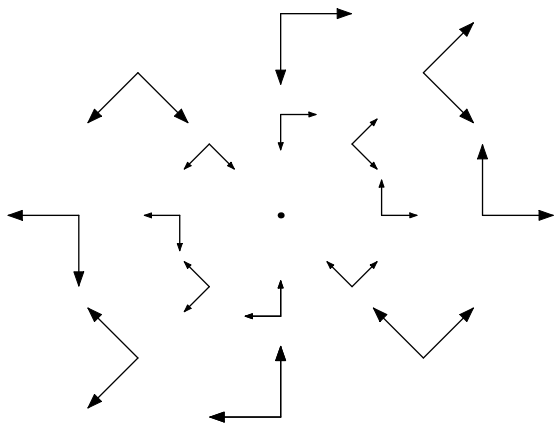}
\caption{This figure shows the orientation and also the strength of
the electric and magnetic field vectors at $t=0$ and in the $z=0$
plane for the solution of the Maxwell equations with a vortex line
described by Eq.(\ref{vortex}). The point in the middle represents
the position of the vortex line. All plots in this paper were
generated with the help of MATHEMATICA \cite{wolfram}.}
\label{fig:1}}
\end{figure}

The simplest solution, the one with a unit vortex strength, is
obtained by choosing $m=2$ in Eq.~(\ref{rs1}). In this case, the
electric and magnetic field vectors are
\begin{subequations}\label{vortex}
\begin{align}
{\bm E}({\bm r},t)&=\sigma B\omega\left(f({\bm r},t),g({\bm r},t),0\right),\\
{\bm B}({\bm r},t)&=\sigma\frac{B\omega}{c}\left(-g({\bm
r},t),f({\bm r},t),0\right),
\end{align}
\end{subequations}
where we introduced the field amplitude expressed in terms of the
magnetic field strength and
\begin{subequations}\label{fg}
\begin{align}
f({\bm r},t)&=x\cos\omega(t-z/c)+\sigma y\sin\omega(t-z/c),\\
g({\bm r},t)&=\sigma x\sin\omega(t-z/c)-y\cos\omega(t-z/c).
\end{align}
\end{subequations}
This simple, linear dependence of the field vectors on $x$ and $y$
will enable us to find explicit solutions of the equations of
motion. The expressions (\ref{vortex}) and (\ref{fg}) show that the
sign factor $\sigma$ determines the helicity --- the sense of
rotation of the pair of vectors ${\bm E}$ and ${\bm B}$. At each
point this pair of vectors rotates with the wave frequency $\omega$
but, in contrast to the circularly polarized plane wave, the
orientation is not the same in the whole plane but the pair rotates
as we move around the vortex line (see Fig.~\ref{fig:1}). In our
case, since the vortex has a unit strength, the full rotation angle
is $2\pi$.

We have shown recently \cite{prl} that there exist analytic
solutions in the classical and quantum theory describing the motion
of charged particles in the presence of the electromagnetic beam
[Eq.\ref{vortex}].  Exact solutions exist because the transverse
motion (in the plane perpendicular to the direction of the beam
propagation) almost separates from the longitudinal motion. The
separation is not complete but the longitudinal motion influences
only the {\em values} of the integration constant appearing in the
equations of motion describing the transverse dynamics. The exact
solvability is not destroyed by the presence of an additional
constant magnetic field aligned with the beam direction. The
presence of this field adds a new dimension to the space of control
parameters and significantly enriches the dynamical behavior of the
system.

The electromagnetic field (\ref{vortex}) may seem unrealistic since
its amplitude increases without bound with the growing distance from
the $z$ axis. However, we shall show that the particle is often
confined to a region close to the $z$ axis and is subject
practically the same field as that of much more realistic
Laguerre-Gaussian or Bessel beams. This expectation has been
recently fully confirmed in \cite{bbc} where we have shown that
numerical solutions of the Lorentz equation in the presence of
Bessel beams near the vortex line exhibit the same behavior as the
analytic solutions found in \cite{prl} for the model wave
(\ref{vortex}).

\section{Classical mechanics}

We shall give first a fully relativistic description of the motion
because the nonrelativistic approximation can be easily obtained
from the relativistic solution by taking the limit $c\to\infty$.

\subsection{Relativistic motion}

The Lorentz equations of motion
\begin{eqnarray}\label{lorentz0}
m\,{\ddot\xi}^{\mu}(\tau)=
e\,f^{\mu\nu}[\xi(\tau)]{\dot\xi}_{\nu}(\tau),
\end{eqnarray}
that govern the motion of a relativistic particle moving in the
presence of the electromagnetic wave [Eq.(\ref{vortex})] and a
constant magnetic field ${\bm B}_0 =(0,0,B_0)$, aligned with the
beam direction, can be written in the form
\begin{subequations}
\label{lorentz}
\begin{align}
{\ddot\xi} &= \sigma\omega_c\,\omega\, f(\xi,\eta,\zeta,\theta)
\left(\dot\theta-\dot\zeta/c\right)+\omega_0{\dot\eta},\label{l1}\\
{\ddot\eta} &= \sigma\omega_c\,\omega\, g(\xi,\eta,\zeta,\theta)
\left(\dot\theta-\dot\zeta/c\right)-\omega_0{\dot\xi},\label{l2}\\
{\ddot\zeta} &= \sigma\frac{\omega_c\,\omega}{c}
[{\dot\xi}f(\xi,\eta,\zeta,\theta)
+{\dot\eta}g(\xi,\eta,\zeta,\theta)],\label{l3}\\
c\,{\ddot\theta}&= \sigma\frac{\omega_c\,\omega}{c}
{\dot\xi}f(\xi,\eta,\zeta,\theta)
+{\dot\eta}g(\xi,\eta,\zeta,\theta)], \label{l4}
\end{align}
\end{subequations}
where $\xi(\tau),\eta(\tau),\zeta(\tau)$ are the space components
and $\theta(\tau)$ is the time component of the four-vector
${\xi}^{\mu}(\tau)$. To save space we dropped the dependence on
$\tau$. The dots denote derivatives with respect to the proper time
$\tau$. The two cyclotron frequencies $\omega_c=eB/m$ and
$\omega_0=eB_0/m$ measure the intensity of the wave and the strength
of the constant magnetic field, respectively. They can be either
positive or negative. The negative value of $\omega_0$ means the
reversal of the magnetic field direction.

From now on, we shall assume that $\sigma=1$ because the equations
of motion (\ref{lorentz}) with the functions $f$ and $g$ defined by
Eq.(\ref{fg}) are invariant under the following transformation
\begin{eqnarray}\label{trans}
\sigma \to -\sigma,\;\;\xi \longleftrightarrow \eta,\;\;\omega_0 \to
-\omega_0.
\end{eqnarray}
Therefore, we do not obtain anything essentially new by considering
both signs of $\sigma$, since to every trajectory for $\sigma=-1$
there corresponds a trajectory for $\sigma=1$ in the reversed
constant magnetic field, differing only by a reflection with respect
to the $x=y$ plane.

The Lorentz equations (\ref{lorentz}) are nonlinear but they can be
effectively linearized owing to conservation laws. In doing so, we
shall follow the procedure developed in Ref.~\cite{prl}. The first
constant of motion $\mathcal E$ is obtained by subtracting
Eq.~(\ref{l3}) from Eq.~(\ref{l4}) which gives
${\ddot\theta}-{\ddot\zeta}/c = 0$ or
${\dot\theta}-{\dot\zeta}/c=\text{const}_1$. Thus, $\mathcal E$ is a
conserved quantity
\begin{eqnarray}\label{const1}
\mathcal{E} = {\dot\theta}-{\dot\zeta}/c =
 \sqrt{1 + ({\dot\xi}^2+{\dot\eta}^2+{\dot\zeta}^2)/c^2}-\dot\zeta/c.
\end{eqnarray}
Apart from the factor $mc^2$, the constant $\mathcal E$ is the
light-front energy --- the conjugate variable to $t_+=t + z/c$
\begin{eqnarray}\label{e}
 \mathcal{E}=\frac{1-v_z/c}{\sqrt{1-{\bm v}^2/c^2}}
 = \frac{\sqrt{m^2c^4 +{\bm p}^2c^2} - p_zc}{mc^2}.
\end{eqnarray}
We shall assume that we start counting the proper time $\tau$ in
such a way that $\theta(0)-\zeta(0)/c=0$. Under this assumption,
Eq.~(\ref{const1}) integrated with respect to $\tau$ yields
\begin{eqnarray}\label{effective}
 \theta - \zeta/c = \mathcal{E}\,\tau.
\end{eqnarray}
Hence, the proper time is simply proportional to the light-front
variable and this will enable us to separate the transverse motion
from the longitudinal motion.

The second constant of motion is obtained by subtracting from
Eq.~(\ref{l3}) the sum of Eq.~(\ref{l1}) multiplied by
$\dot\xi/\mathcal E$ and Eq.~(\ref{l2}) multiplied by
$\dot\eta/\mathcal E$. Since the right hand sides cancel, we obtain
\begin{eqnarray}\label{const2}
{\dot\zeta} - \frac{1}{2c\mathcal{E}}({\dot\xi}^2+{\dot\eta}^2) =
\text{const}_2 = \frac{c}{2}(\frac{1}{\mathcal{E}}-\mathcal{E}).
\end{eqnarray}
The last expression is obtained by rearranging the formula $1 +
({\dot\xi}^2+{\dot\eta}^2+{\dot\zeta}^2)/c^2 =(\mathcal
E+\dot\zeta/c)^2$ obtained from Eq.~(\ref{const1}).

Owing to Eq.~(\ref{const1}), the longitudinal motion affects the
transverse motion only through the value of $\mathcal E$ and since
it is a constant of motion the equations for $\xi$ and $\eta$ may be
solved first. The integration of Eqs.~(\ref{l1}) and (\ref{l2}) is
made simpler by introducing a complex combination of the coordinates
$Z=\xi+i\eta$. The equation of motion for $Z$ obtained from
Eqs.~(\ref{lorentz}) reads
\begin{eqnarray}\label{weq}
{\ddot Z}= \omega_c\Omega Z^* e^{i\Omega\tau}-i\omega_0 \dot Z,
\end{eqnarray}
where in addition to the two cyclotron frequencies $\omega_c$ and
$\omega_0$ associated with the strength of the wave and the constant
magnetic field we defined the third frequency $\Omega$ --- an
effective wave frequency as seen by the moving particle. Thus, the
dynamics of the particle is governed by the three frequencies
\begin{eqnarray}\label{def}
\Omega = \omega\mathcal{E},\;\;\omega_c = \frac{e B}{m},\;\;\omega_0
= \frac{e B_0}{m}.
\end{eqnarray}
The form ${\ddot Z}e^{-i\Omega\tau/2}= \omega_c\Omega(Z
e^{-i\Omega\tau/2})^*-i\omega_0 \dot Z e^{-i\Omega\tau/2}$ of
Eq.~(\ref{weq}) suggests a transformation to the frame rotating (in
proper time) with the angular velocity $\Omega/2$ around the
$z$-axis. This amounts to replacing $\xi$ and $\eta$ by the new
variables
\begin{subequations}\label{new}
\begin{align}
\alpha(\tau) &= \xi(\tau)\cos(\Omega\tau/2) + \eta(\tau)\sin(\Omega\tau/2),\\
\beta(\tau) &= -\xi(\tau)\sin(\Omega\tau/2) +
\eta(\tau)\cos(\Omega\tau/2),
\end{align}
\end{subequations}
or in complex notation
\begin{eqnarray}\label{wdef}
Z=\xi+i\eta=(\alpha+i\beta)e^{i\Omega\tau/2}.
\end{eqnarray}
Upon substituting this expression into Eq.~(\ref{weq}), we get rid
of an explicit $\tau$ dependence and obtain the following equations
with constant coefficients for $\alpha $ and $\beta$
\begin{subequations}\label{aeq}
\begin{align}
{\ddot\alpha} &= (\Omega^2/4+\Omega\omega_0/2+\Omega\omega_c)\alpha + (\Omega+\omega_0){\dot\beta},\\
{\ddot\beta} &= (\Omega^2/4+\Omega\omega_0/2-\Omega\omega_c)\beta
-(\Omega+\omega_0){\dot\alpha}.
\end{align}
\end{subequations}
The mutual relationships between the three frequencies $\Omega$,
$\omega_c$, and $\omega_0$ determine whether the transverse motion
is localized near the $z$ axis or is unbounded.

A full description is most easily done in the Hamiltonian formalism.
The equations of motion (\ref{aeq}) can be obtained from the
following Hamiltonian
\begin{eqnarray}\label{ham}
H = \frac{p_\alpha^2 + p_\beta^2}{2m}
+ m \frac{(a-b)\alpha^2+(a+b)\beta^2}{2} \nonumber\\
- w(\alpha p_\beta - \beta p_\alpha),\hspace{2cm}
\end{eqnarray}
where
\begin{eqnarray}\label{fgw}
a = \frac{\omega_0^2}{4},\;\; b = \Omega\omega_c,\;\; w
=\frac{\Omega+\omega_0}{2}.
\end{eqnarray}
This Hamiltonian (for $\omega_0=0$) differs by a canonical
transformation from the one considered in Ref.~\cite{prl}. We found
this new form of the Hamiltonian more convenient than the previous
one. The canonical equations of motion can be written in the
following matrix form
\begin{eqnarray}\label{hameq}
i\frac{d}{dt}X = X\!\cdot\!{\hat M},
\end{eqnarray}
where
\begin{eqnarray}\label{x}
X = \left(
\begin{array}{cccc}
\alpha, & \beta, & p_\alpha, & p_\beta
\end{array}\right),
\end{eqnarray}
and
\begin{eqnarray}\label{m}
{\hat M} = \left(
\begin{array}{cccc}
0&-iw&-im(a-b)&0\\
iw&0&0&-im(a+b)\\
i/m&0&0&-iw\\
0&i/m&iw&0
\end{array}\right).
\end{eqnarray}
We have absorbed $i$ into the definition of the matrix ${\hat M}$ to
make its eigenvalues equal to the (real) characteristic frequencies.
These are the same equations of motion (only with different values
of the coefficients) that govern the dynamics of particles in the
mechanical model of the Paul trap \cite{paul,bbs1}, Trojan asteroids
in the Sun-Jupiter system, and Trojan wave packets of Rydberg
electrons in an electromagnetic wave \cite{bke} or in a molecule
with an electric dipole moment \cite{bbbb}. As a matter of fact,
every quadratic Hamiltonian, by an appropriate linear canonical
transformation, can be reduced to the form (\ref{ham}).

The characteristic frequencies in the present case are
\begin{align}\label{charf}
\Omega_\pm &= \Omega r_{\pm}\nonumber\\
 r_{\pm} &= \frac{1}{2}\sqrt{(1+\mu)^2+\mu^2\pm 4\sqrt{\nu^2+\mu^2(1+\mu)^2/4}}\;,
\end{align}
where we used $\Omega$ as a yardstick to measure all frequencies,
i.e.
\begin{eqnarray}\label{par}
\mu=\omega_0/\Omega,\;\;\nu=\omega_c/\Omega.
\end{eqnarray}
The description of the motion with the use of two dimensionless
parameters $\mu$ and $\nu$ is very convenient because it exhibits a
self-similarity inherent in our problem. The dimensionless ratios
$\mu$ and $\nu$ provide a unified description of different physical
situations. For example, a particle in a 2.45\,GHz microwave oven
will show the same behavior as a particle in an optical wave of a
505\,nm blue laser provided we increase the amplitude of the wave
and the strength of the magnetic field by a factor of $2.42 10^5$.
An increase or decrease of all three frequencies by the same factor
does not change the character of the motion but it results in a
decrease or increase of characteristic space and time scales. We
shall take this fact into account and express the distances in units
of the angular wavelength $\lambdabar=c/\omega$ and the time in
units of an angular period $1/\omega$. These units of length and
time are used in all the figures in this paper. In other words, we
shall use the units in which $\omega=1$ and $c=1$.

Bounded oscillations occur only when both characteristic frequencies
$\Omega_\pm$ are real and this means that
\begin{eqnarray}\label{cond}
\vert\nu\vert<\frac{1}{2}\left\vert\frac{1}{2}+\mu\right\vert.
\end{eqnarray}

\begin{figure}[ht]
\centering {\includegraphics[width=0.5\textwidth]{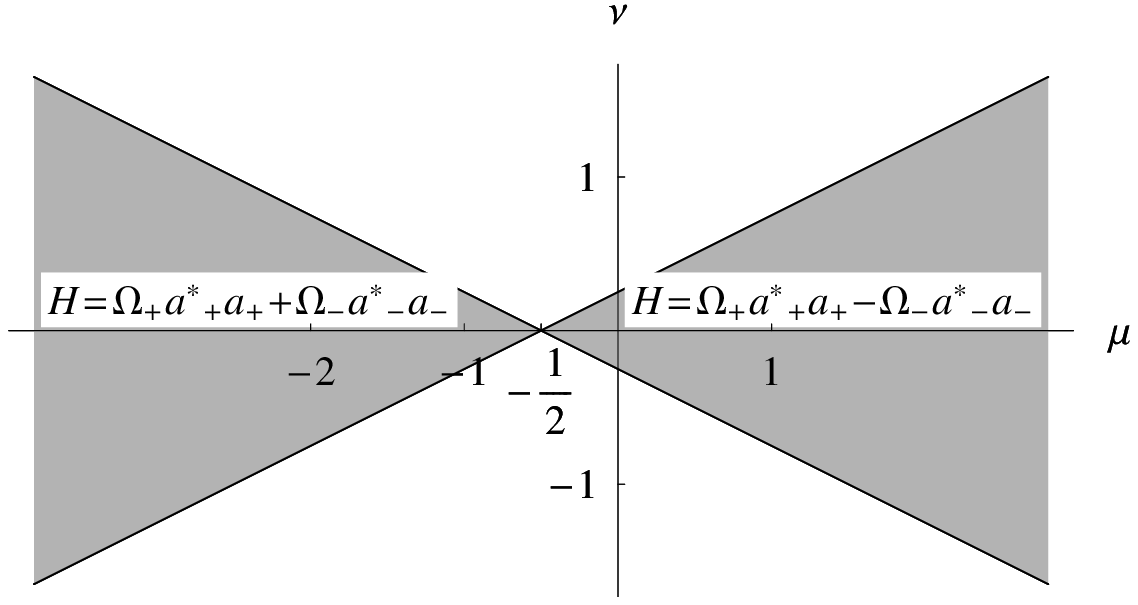}
\caption{Regions of stability in the $\mu\nu$ plane are shown as two
shaded areas. In the area on the left hand side we have a normal
harmonic oscillator while in the area on the right hand side we have
the Trojan regime --- stable oscillations but the Hamiltonian is not
positive definite.} \label{fig:2}}
\end{figure}

The mode amplitudes $a_\pm$ and $a^*_\pm$ (classical counterparts of
the annihilation and creation operators) are found by solving the
eigenvalue problem for the matrix ${\hat M}$ appearing in
Eq.~(\ref{hameq})
\begin{subequations}\label{annih}
\begin{eqnarray}
 a_+ = \frac{1}{N_+^{1/2}}
\Big[t_{++}\left(p_\beta-\frac{m\Omega s_-}{1+\mu}\alpha\right)\;\;\;\nonumber\\
- i \,\varepsilon \,t_{+-}\left(p_\alpha+\frac{m\Omega s_+}{1+\mu}\beta\right)\Big],\\
a_- = \frac{1}{N_-^{1/2}}
\Big[t_{--}\left(p_\beta+\frac{m\Omega s_+}{1+\mu}\alpha\right)\;\;\;\nonumber\\
+ i\varepsilon \,t_{-+}\left(p_\alpha-\frac{m\Omega
s_-}{1+\mu}\beta\right)\Big],\;\;
\end{eqnarray}
\end{subequations}
where to save space we introduced the following functions of $\mu$
and $\nu$
\begin{subequations}\label{rsdef}
\begin{align}
s_\pm &= \sqrt{\nu^2+\mu^2(1+\mu)^2/4}\pm\nu,\\
t_{+\pm}&=\sqrt{(1+\mu)^2+ 2s_\pm},\\
t_{-\pm}&=\sqrt{\vert(1+\mu)^2-2s_\pm\vert},\\
\varepsilon &={\rm sgn}(1+\mu),
\end{align}
\end{subequations}
and the normalization factors are
\begin{eqnarray}\label{norm}
N_\pm = 4m\Omega_\pm(s_++s_-).
\end{eqnarray}
This normalization guarantees that $a_\pm$ and $a^*_\pm$ have the
canonical Poisson brackets
\begin{eqnarray}\label{pb}
\{a_\pm,a^*_\pm\}=-i.
\end{eqnarray}
The formulas (\ref{annih}) are not valid when $\mu=-1$. However, in
this special case the equations of motion (\ref{aeq}) describe just
two uncoupled harmonic oscillators and the determination of
eigenmodes is very simple. The Hamiltonian [Eq.(\ref{ham})]
expressed through the amplitudes $a_\pm$ reads
\begin{eqnarray}\label{ham1}
H = \Omega_+\,a_+^*a_+ - {\rm sgn}(1/2+\mu)\Omega_-\,a_-^*a_-.
\end{eqnarray}
The minus sign in the diagonal form of the Hamiltonian, indicates
that for $1/2+\mu>0$ the oscillations in the transverse plane are
bounded due to the Coriolis force --- a characteristic feature of
the Paul trap or Trojan asteroids and Trojan electrons. We would
like to stress that despite its quadratic form the Hamiltonian of
the transverse motion does not correspond to a simple harmonic
oscillator in two dimensions because the frequencies of oscillations
$\Omega_\pm$ are not fixed once and for all but they depend on the
initial velocities through the parameter $\mathcal E$. The regions
of stable motion in the $\mu\nu$ plane are shown in
Fig.~\ref{fig:2}.

The general solution for $\xi(\tau)$ and $\eta(\tau)$ in the
laboratory frame is obtained by solving Eq.~(\ref{aeq}) in terms of
eigenmodes and then undoing the rotation in Eq.(\ref{new}). The
final expression for the motion of particles in the plane
perpendicular to the vortex line can be compactly written in the
same complex form as in \cite{prl}
\begin{eqnarray}\label{sol}
\xi(\tau)+i\eta(\tau)=
e^{i\Omega\tau/2}\big[(iD\kappa_++A)\sin(\Omega_+\tau)\nonumber\\
-(B\kappa_-+iC)\sin(\Omega_-\tau)
+(iA\kappa_+-D)\cos(\Omega_+\tau)\nonumber\\
+(C\kappa_--iB)\cos(\Omega_-\tau)\big],
\end{eqnarray}
but the meaning of $A,B,C,D$, and $\kappa_\pm$ is now different,
namely
\begin{equation}\label{kappa}
\kappa_\pm=\frac{r_\pm^2+1/4+\mu/2\pm\nu}{(1+\mu)r_\pm},
\end{equation}
and the constants $A,B,C$, and $D$ are the following functions of
the initial values of the transverse positions and velocities
\begin{subequations}\label{sol1}
\begin{align}
A&=\frac{(1/2-\kappa_-r_-)\eta_0+{\dot\xi}_0/\Omega}
{r_+-r_-\kappa_-\kappa_+},\\
B&=\frac{(\kappa_+/2-r_+)\eta_0
+\kappa_+{\dot\xi}_0/\Omega}{r_+-r_-\kappa_-\kappa_+},\\
C&=\frac{(1/2-\kappa_+r_+)\xi_0-{\dot\eta}_0/\Omega}
{r_--r_+\kappa_-\kappa_+},\\
D&=\frac{(\kappa_-/2-r_-)\xi_0-\kappa_-{\dot\eta}_0/\Omega}
{r_--r_+\kappa_-\kappa_+}.
\end{align}
\end{subequations}

Having found the solution in the transverse plane, we can now just
integrate Eq.~(\ref{const2}) to obtain the following formula for the
longitudinal motion
\begin{eqnarray}\label{zeta0}
\zeta(\tau) =
\frac{c\tau}{2}(\frac{1}{\mathcal{E}}-\mathcal{E})+\int_0^\tau
d\tau\frac{1}{2c\mathcal{E}}({\dot\xi}^2+{\dot\eta}^2)+\zeta(0).
\end{eqnarray}
The integration is cumbersome but elementary and it leads to
\begin{widetext}
\begin{eqnarray}\label{zeta}
\zeta(\tau)=\frac{c\tau}{2}\Bigg[\frac{1}{\cal E}-{\cal
E}+\frac{(A^2+D^2)\left((\Omega^2+4\Omega_+^2)
(1+\kappa_+^2)-8\Omega\Omega_+\kappa_+\right)+(B^2+C^2)
\left((\Omega^2+4\Omega_-^2)(1+\kappa_-^2)-8\Omega\Omega_-\kappa_-\right)} {8c^2{\cal E}}\Bigg]\nonumber\\
+\frac{(1-\kappa_+^2)(\Omega^2-4\Omega_+^2)
\left[(D^2-A^2)\sin(2\Omega_+\tau)-2AD(1-\cos(2\Omega_+\tau))\right]}{32c{\cal E}\Omega_+}\nonumber\\
+\frac{(1-\kappa_-^2)(\Omega^2-4\Omega_-^2)
\left[(B^2-C^2)\sin(2\Omega_-\tau)
+ 2BC(1-\cos(2\Omega_-\tau))\right]}{32c{\cal E}\Omega_-} \nonumber\\
+\frac{(\kappa_+-\kappa_-)(\Omega^2-4\Omega_+\Omega_-)
+2(\kappa_+\kappa_--1)\Omega\Omega_m}{8c{\cal E}\Omega_p}
\bigg[(CD-AB)\sin(\Omega_p\tau) -(AC+BD)(1-\cos(\Omega_p\tau))\bigg]\nonumber\\
-\frac{(\kappa_++\kappa_-)(\Omega^2+4\Omega_+\Omega_-)
-2(\kappa_+\kappa_-+1)\Omega\Omega_p}{8c{\cal E}\Omega_m}
\bigg[(CD+AB)\sin(\Omega_m\tau)-(AC-BD)
(1-\cos(\Omega_m\tau))\bigg]+\zeta(0),\hspace{.4cm}
\end{eqnarray}
\end{widetext}
where $\Omega_p=\Omega_++\Omega_-$ and $\Omega_m=\Omega_+-\Omega_-$.
Thus, the motion in the $z$ direction is a composition of a uniform
motion in the proper time $\tau$ (not in the laboratory time $t$)
and oscillations with four frequencies $2\Omega_\pm$ and
$\Omega_+\pm\Omega_-$. By a fine tuning of the initial conditions we
may cancel the uniform motion completely and leave only the
oscillations, but any departure from these special values will cause
a uniform drift. The appearance of two additional frequencies
$\Omega_+\pm\Omega_-$ is a new feature that is not found in the
absence of the constant magnetic field. Note that the motion in the
transverse plane is not completely decoupled from the longitudinal
motion because the effective frequency $\Omega$, appearing
abundantly in the formula (\ref{sol}), depends (through $\mathcal
E$) on the initial velocity in the longitudinal direction.

\begin{figure}[ht]
\centering {\includegraphics[width=0.4\textwidth]{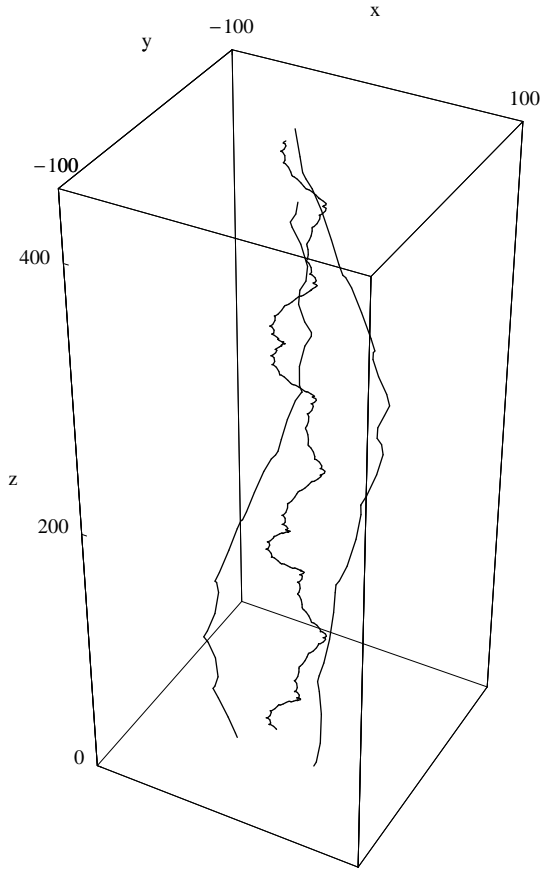}
\caption{Three trajectories obtained for $\mu=0.075$, $\nu=0.1$, and
${\bm v}/c=(0.001,0,0)$, differing only in the initial
positions.}\label{fig:3}}
\end{figure}

The presence of stability and instability windows is a manifestation
of the {\em parametric resonance}. In the region of instability the
amplitude of oscillations grows exponentially. This is essentially
different as compared to the motion of a charged particle in the
circularly polarized plane wave and a constant magnetic field. In
that case there appears a standard resonance, i.e. the amplitude
grows linearly when the frequency of the driving force (the wave)
exactly matches the characteristic frequency of the system
(cyclotron frequency). When the circularly polarized wave is
replaced by a wave with a vortex, the set of linear {\em
inhomogeneous} differential equations becomes a set of {\em
homogeneous} equations with periodically varying coefficients. The
driving force disappears but the coefficients of the equation become
now time dependent and there appears a parametric resonance. Its
characteristic features are: The appearance of the whole regions of
instability (and not only discrete values as in the case of an
ordinary resonance), which shrink to just one point ($\omega_0 =
-\omega/2$) when $\omega_c\rightarrow 0$ and an exponential growth
of the amplitude in all regions of instability.

\begin{figure*}[ht]
\centering {\includegraphics[width=14cm]{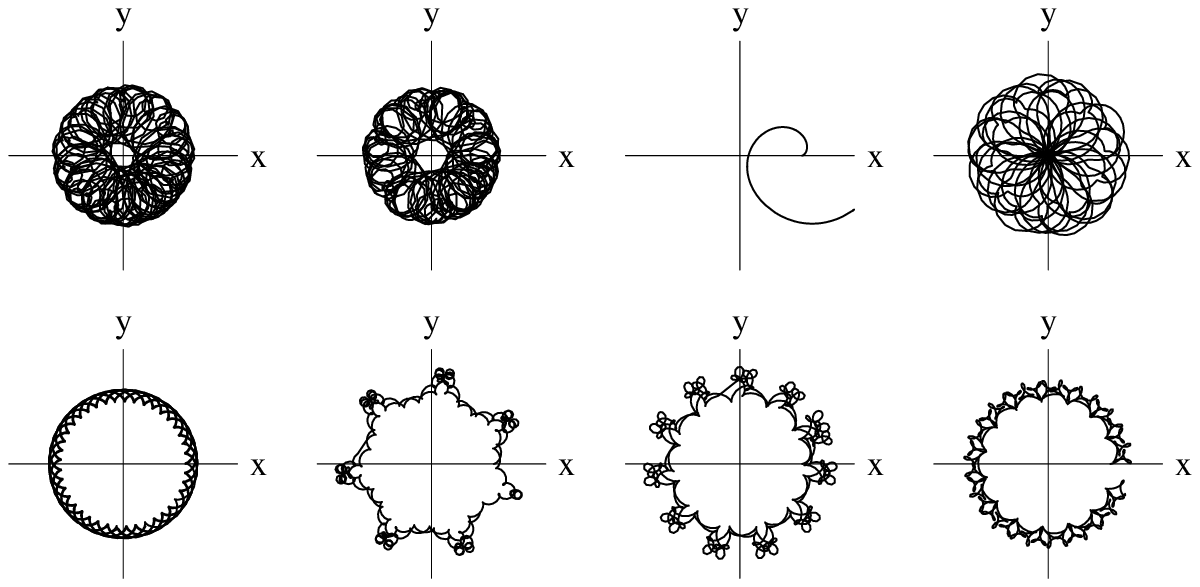} \caption{The
trajectories of a particle projected on the $xy$ plane are shown for
different values of the parameter $\mu$ (i.e. for different values
of the constant magnetic field) $\mu$ = (-0.75, -0.7, -0.5, -0.2,
0.0, 0.1, 0.2, 0.4). The value of $\nu$ has been fixed at
$\nu=0.075$. The initial values of the position vectors (measured in
$\lambdabar=c/\omega$) and velocity vectors (measured in $c$) are
$(60,0,0)$ and $(0.001,\,0,\,0.001)$, respectively. The size of the
area shown in these plots is $100\lambdabar\times 100\lambdabar$ and
the time lapse is $600/\omega$. The third plot shows the exponential
growth of the distance from the vortex line characteristic of the
unstable regime.} \label{fig:4}}
\end{figure*}

\begin{figure*}[ht]
\centering {\includegraphics[width=14cm]{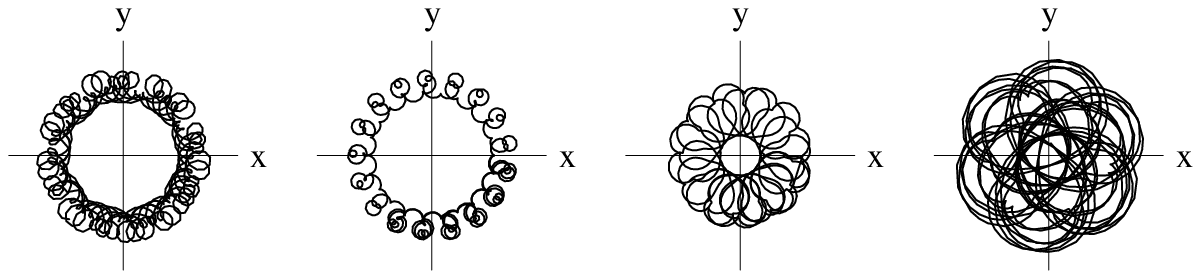} \caption{The
trajectories of a particle obtained for the same initial conditions
as in Fig.~\ref{fig:4}. In these plots the value of $\mu$ is fixed
at $\mu=-0.2$ but the strength of the field amplitude varies as
follows: $\nu = (-0.1,\,-0.05,\,0.05,\,0.1)$.} \label{fig:5}}
\end{figure*}

\begin{figure*}[ht]
\centering {\includegraphics[width=14cm]{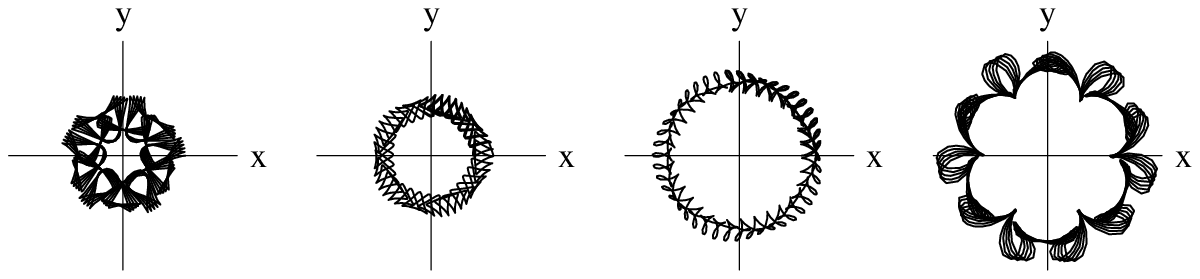} \caption{The
trajectories of a particle obtained for the same initial conditions
as in Fig.~\ref{fig:4}. In these plots the value of $\mu$ is fixed
at $\mu=0.5$ but the strength of the field amplitude varies as
follows $\nu = (-0.2,\,-0.1,\,0.1,\,0.2)$.} \label{fig:6}}
\end{figure*}
Three typical stable three-dimensional relativistic trajectories
differing by the choice of initial positions are shown in
Fig.~\ref{fig:3}. The trajectories of a particle projected on the
$xy$ plane for different values of the parameters $\mu$ and $\nu$
are shown in Figs.~\ref{fig:4}--\ref{fig:6}. The plots demonstrate
that the characteristic shapes of the trajectories are very
sensitive to the changes of these parameters.

\subsection{Nonrelativistic limit}

The solutions of the equations of motion in the nonrelativistic
regime can be obtained from the relativistic ones just by formally
taking the limit $c\to\infty$. In this limit, the difference between
the proper time and the laboratory time disappears. Also, the wave
frequency $\omega$ and the effective frequency $\Omega$  become
equal, because ${\mathcal E}=1$. The motion in the longitudinal
direction completely decouples from the oscillations in the
transverse plane. In the nonrelativistic limit $c(1/{\mathcal
E}-{\mathcal E})/2\to v_{z}$. In this limit, the motion in the $z$
direction becomes a free motion described by the formula
$z(t)=z_0+v_{z}t$. In contrast to the relativistic case, the
velocity in the $z$ direction does not show any oscillations,
$v_z=v_{z0}$. We arrive at the same conclusions by solving the
nonrelativistic equations of motion ($\sigma=1$)
\begin{subequations}
\label{newton}
\begin{align}
\frac{d^2\xi}{dt^2} &= \omega_c\,\omega\,(\xi\cos\omega t +
\eta\sin\omega t)
+\omega_0\frac{d\eta}{dt},\label{n1}\\
\frac{d^2\eta}{dt^2} &= \omega_c\,\omega\,(\xi\sin\omega t
-\eta\cos\omega t)
-\omega_0\frac{d\xi}{dt},\label{n2}\\
\frac{d^2\zeta}{dt^2} &= 0, \label{n3}
\end{align}
\end{subequations}
which are obtained from the Lorentz equations (\ref{lorentz}) in the
limit $c\to \infty$. Since the decoupling of the longitudinal and
the transverse motion in the nonrelativistic limit is complete, we
may easily obtain in this case also the solutions for a
two-dimensional gas of mutually noninteracting charged particles
kept near the surface by an additional confining potential $V(z)$.

\begin{figure}[ht]
\centering {\includegraphics[width=0.50\textwidth]{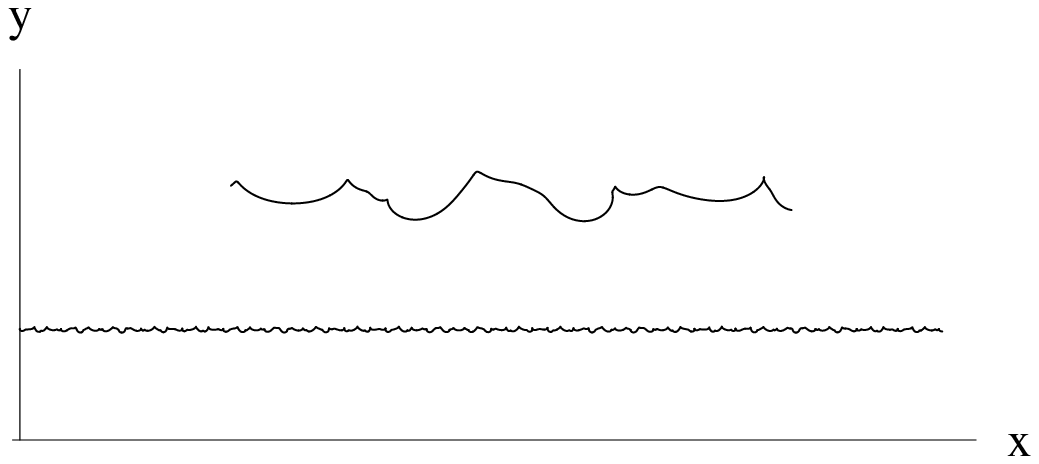}
\caption{The transport of an orbit by a moving vortex line. On top
of the average motion with a constant velocity in the $x$ direction
there are additional oscillations, clearly seen after a tenfold
enlargement in the upper part of the figure.} \label{fig:7}}
\end{figure}
\begin{figure}[ht]
\centering {\includegraphics[width=0.5\textwidth]{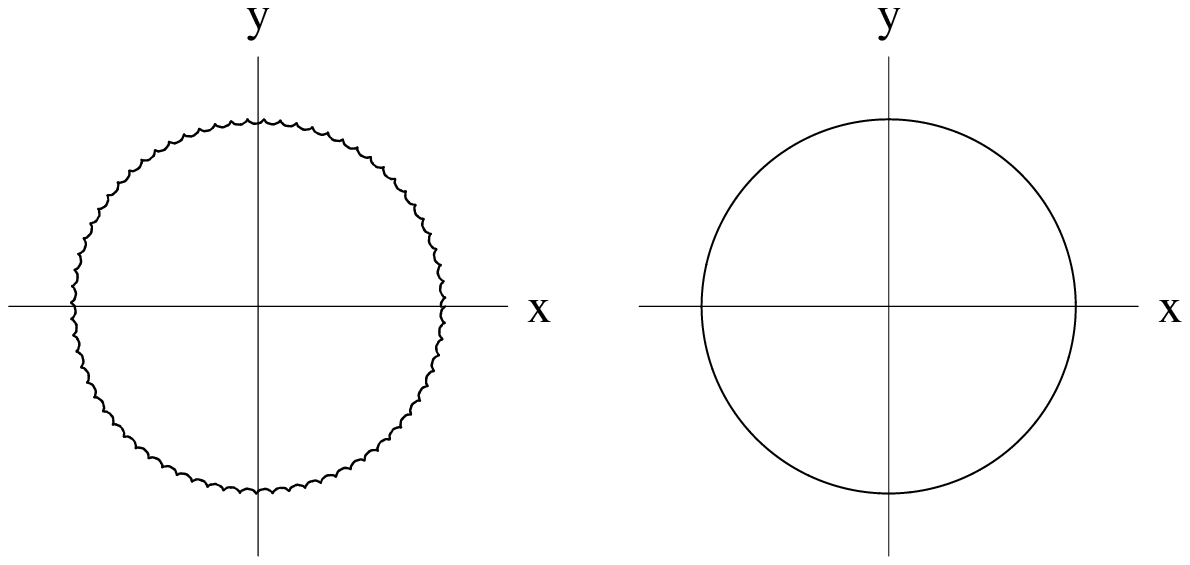}
\caption{The transport of an orbit by a vortex line sweeping out the
surface of a cylinder with the radius $10\,\lambdabar$. The field
parameters are $\mu=0.1$ and $\nu=0.12$. The actual particle
trajectory (upper plot) is compared with the path that traces the
points of the intersection of the vortex line with the $xy$ plane
defined by the equation $z=\zeta[\tau]$ (lower plot). The period of
the circular motion was chosen as 1000 times the period $T$ of the
carrier wave and the time lapse for the trajectory shown in this
figure is equal to four full periods of the circular motion. The
distances are measured in units of the wavelength
$\lambdabar=c/\omega$ of the carrier wave. Initially the particle
was placed at the position of the vortex ($\xi_0=10,\,\eta_0=0$)
with zero velocity. A slow drift in the $z$ direction is due to
relativistic effects.}\label{fig:8}}
\end{figure}

\begin{figure}[ht]
\centering {\includegraphics[width=0.5\textwidth]{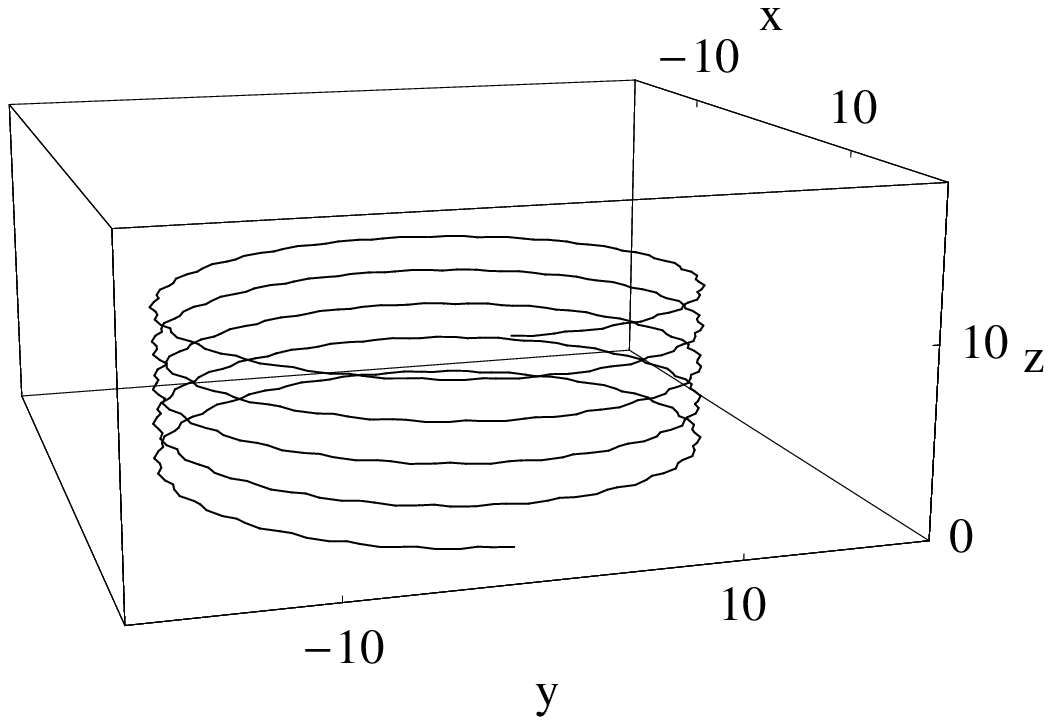}
\caption{The transport of an orbit by a vortex line sweeping out a
cylindrical surface with the cross section in the form of a trefoil.
This figure shows only the projection of the trajectory on the $xy$
plane. The two frequencies in the formula (\ref{rs3}) are
$1.0005\,\omega$ and $0.999\,\omega$. Since the two detunings are
commensurate, the trajectory (right) of the vortex line in the $xy$
plane is closed. The parameters characterizing the position of the
vortex are $x_1=5,\,y_1=0,\,x_2=0$, and $y_2=-10$. The remaining
parameters are $\mu=10$ and $\nu=1.5$. The distances are measured in
units of the wavelength $\lambdabar=c/\omega$ of the carrier wave.
Initially the particle is placed on the vortex line at $\xi_0=5$ and
$\eta_0=-10$ with zero velocity. The size of the area in these
figures is $20\lambdabar\times 20\lambdabar$.}\label{fig:9}}
\end{figure}

\section{Transport of orbits by moving vortices}

Orbits trapped by vortices can be moved around across the magnetic
field. The simplest case is a uniform motion. The electromagnetic
field with a moving vortex line can be easily obtained from the
field with a stationary vortex line [Eq.(\ref{vortex})] by a Lorentz
transformation. Assuming that the motion is along the $x$ axis with
velocity $v$, we obtain the following transformed electric and
magnetic field vectors
\begin{subequations}\label{vortexl}
\begin{eqnarray}
\frac{{\bm E}_v}{B\omega}= \left(
\begin{array}{c}
\gamma(x-vt)\cos\varphi+ y\sin\varphi\\
\gamma^2(x-vt)\sin\varphi-\gamma y\cos\varphi\\
-v(\gamma^2(x-vt)\cos\varphi+\gamma y\sin\varphi)/c
\end{array}
\right),\quad\\
\frac{c{\bm B}_v}{B\omega}= \left(\!
\begin{array}{c}
-\gamma(x-vt)\sin\varphi+ y\cos\varphi\\
\gamma^2(x-vt)\cos\varphi+\gamma y\sin\varphi\\
v(\gamma^2(x-vt)\sin\varphi-\gamma y\cos\varphi)/c
\end{array}
\right),\quad
\end{eqnarray}
\end{subequations}
where $\gamma=1/\sqrt{1-v^2/c^2}$ and $\varphi =
\omega[\gamma(t-vx/c^2)-z/c]$. In Fig.~\ref{fig:7} we show the
trajectory of a particle that is being pulled across the constant
magnetic field by the uniformly moving vortex line. Note that this
is not the same as transforming the whole problem to a moving frame
by a Lorentz transformation. Such a transformation would result also
in a change of the constant magnetic field into a crossed magnetic
and electric fields.

The transport with a uniform average velocity is just the simplest
example, but there is a plethora of more intricate cases, all of
them exhibiting the same behavior. The simplest method to construct
a solution of the Maxwell equations with a vortex line undergoing
complicated motions is to superimpose our basic solution
[Eq.(\ref{vortex})] with circularly polarized plane waves moving in
the same direction, slightly detuned from the frequency $\omega$. We
shall consider here two cases: a superposition with just one plane
wave and a superposition with two waves. In the first case, the
vortex position at a fixed value of $z$ will move on a circle and in
the second case it will follow a trefoil figure.

In the circular case, we shall choose the Riemann-Silberstein vector
in the form
\begin{align}\label{rs2}
&{\bm F}_c({\bm r},t) = B\omega\left({\hat{\bf x}}+i{\hat{\bf y}}\right)\nonumber\\
&\times\left[(x+iy)e^{-i\omega(t-z/c)} -
(x_1+iy_1)e^{-i\omega_1(t-z/c)}\right],
\end{align}
where $x_1$ and $y_1$ are two parameters that fix the radius of the
circle and the initial position of the vortex. The vanishing of the
RS vector determines the position $(x_v,y_v)$ of the vortex line as
a function of $z$ and $t$
\begin{eqnarray}\label{pos1}
x_v(z,t)+iy_v(z,t) = (x_1+iy_1)e^{i(\omega-\omega_1)(t-z/c)}.
\end{eqnarray}
The vortex forms a screw line of the radius $\sqrt{x_1^2+y_1^2}$ and
the step $2\pi c/(\omega-\omega_1)$ rotating with the angular
frequency $\omega-\omega_1$ around the $z$ axis. The formulas for
the electric and magnetic field vectors read
\begin{subequations}\label{vortexc}
\begin{eqnarray}\label{vortexca}
\frac{{\bm E}_c}{B\omega}=\!\left(\!\begin{array}{c}
x\cos\vartheta+ y\sin\vartheta-x_1\cos\vartheta_1-y_1\sin\vartheta_1\\
x\sin\vartheta-y\cos\vartheta-x_1\sin\vartheta_1+y_1\cos\vartheta_1\\
0
\end{array}
\!\!\right),\quad\\
\frac{c{\bm B}_c}{B\omega}=\!\left(\!\begin{array}{c}
y\cos\vartheta-x\sin\vartheta-y_1\cos\vartheta_1+x_1\sin\vartheta_1\\
y\sin\vartheta+x\cos\vartheta-y_1\sin\vartheta_1-x_1\cos\vartheta_1\\
0
\end{array}\!
\!\!\right)\!,\quad
\end{eqnarray}
\end{subequations}
where $\vartheta=\omega(t-z/c)$ and $\vartheta_1=\omega_1(t-z/c)$.
Thus, the vortex line forms a helix rotating with the angular
frequency $\omega-\omega_1$. The projection of this helix on the
$xy$ plane forms a circle. In Fig.~\ref{fig:8} we show the motion of
the particle dragged by such a helical vortex line. The values of
the dimensionless parameters $\mu=0.1$ and $\nu=0.12$ that determine
the character of the motion may correspond to different field and
particle configurations. For example, they can describe an electron
moving in a constant magnetic field of 1 T and in an electromagnetic
microwave of the frequency $2.8 10^{11}$Hz and characteristic
intensity $3.4 10^{10}{\rm W/cm}^2$. These parameters are not chosen
to be very realistic (for example, the intensity is very high) but
they were selected to enhance characteristic features of our
solution.

In order to obtain a more elaborate path (a generalized Lissajou
figure) of the vortex in the $xy$ plane, we add two plane waves
instead of one, i.e.
\begin{align}\label{rs3}
{\bm F}_g({\bm r},t) = B\omega\left({\hat{\bf x}}+i{\hat{\bf
y}}\right) &\big[(x+iy)e^{-i\omega(t-z/c)}\nonumber\\
-(x_1+iy_1)e^{-i\omega_1(t-z/c)}&-(x_2+iy_2)e^{-i\omega_2(t-z/c)}\big].
\end{align}
In this case, the position of the vortex as a function of $t$ and
$z$ is given by the equation
\begin{eqnarray}\label{pos2}
x_v(z,t)+iy_v(z,t) &= (x_1+iy_1)e^{i(\omega-\omega_1)(t-z/c)}\nonumber\\
&+(x_2+iy_2)e^{i(\omega-\omega_2)(t-z/c)}.
\end{eqnarray}

The numerical solutions presented here in
Figs.~\ref{fig:4}--\ref{fig:9} should leave no doubt that an
electromagnetic vortex may easily overcome the inertia of the
particle and drag the cyclotron orbit along. Analytic solutions of
the equations of motion for a relativistic particle in the presence
of electromagnetic waves with moving vortices can also be obtained
and they will be described elsewhere.

\section{Quantum mechanics}

We shall first describe the motion of a quantum particle in the
combination of an electromagnetic wave with a vortex line
[Eq.(\ref{vortex})] and a constant magnetic field in a fully
relativistic case. The nonrelativistic limit will be obtained by a
simple reinterpretation of the parameters. Also, the incorporation
of spin does not introduce any significant complications. Therefore,
we shall start with the Dirac equation.

\subsection{Relativistic quantum mechanics}

The Dirac equation for a particle moving in the presence of an
electromagnetic field described by a vector potential ${\bm A}({\bm
r},t)$ can be written in the form
\begin{eqnarray}\label{dirac0}
i\hbar\partial_t\Psi = \left[c{\bm\alpha}\!\cdot\!(-i\hbar{\bm
\nabla} - e{\bm A}) + \beta mc^2\right]\Psi.
\end{eqnarray}
In our case, the vector potential can be chosen as
\begin{equation}\label{pot}
{\bm A}({\bm r},t)=\left(
\begin{array}{c}
-B g({\bm r},t)-yB_0/2\\
B f({\bm r},t)+xB_0/2\\
0\end{array}\right).
\end{equation}
The quantum counterpart of the classical symmetry transformation
(\ref{trans}), namely
\begin{eqnarray}\label{trans1}
\sigma \to -\sigma,\;x \longleftrightarrow y,\;\omega_0 \to
-\omega_0,\; \Psi \to \frac{1+\alpha_x\alpha_y}{\sqrt{2}}\Psi,
\end{eqnarray}
leaves the Dirac equation invariant. Therefore, we may again
restrict ourselves to the case $\sigma=1$ and obtain the solution
for $\sigma=-1$ by applying the transformation (\ref{trans1}). It is
convenient to choose the Dirac representation \cite{dirac} of
$\alpha_i$ and $\beta$ but with the matrices $\alpha_z$ and $\beta$
interchanged, i.e.
\begin{eqnarray}\label{dirmat}
\alpha_x=\rho_x\sigma_x,\;\alpha_y=\rho_x\sigma_y,\;\alpha_z=\rho_z,\;\beta=\rho_x\sigma_z,
\end{eqnarray}
where $\rho_i$ and $\sigma_i$ are two sets of Pauli matrices acting
independently. Next, we split the four component bispinor into
two-component wave functions with the help of two projection
operators $P_\pm=(1\pm\alpha_z)/2$. In the representation
(\ref{dirmat}), the matrix $\alpha_z$ is diagonal and the splitting
amounts to taking the upper and lower components of the Dirac
bispinor $\Psi$. Therefore, we may write
\begin{eqnarray}\label{proj}
\Psi=P_+\Psi + P_-\Psi = \left(\begin{array}{c}\Psi_+\\0
\end{array}
\right) + \left(\begin{array}{c} 0\\\Psi_-
\end{array}
\right)
\end{eqnarray}
and the Dirac equation (\ref{dirac0}) takes on the form
\begin{subequations}\label{dirac1}
\begin{eqnarray}
2i\hbar\partial_+\Psi_+ = {\hat h}\,\Psi_-,\label{dirac1a}\\
2i\hbar\partial_-\Psi_- = {\hat h}\,\Psi_+,\label{dirac1b}
\end{eqnarray}
\end{subequations}
where $\partial_\pm=\partial/\partial{t_\pm}=(\partial_t\pm
c\partial_z)/2$ and
\begin{eqnarray}\label{hamdir}
{\hat h} =
c\left[mc\sigma_z-{\bm\sigma}_\perp\!\cdot\!(i\hbar\nabla+e{\bm
A})\right].
\end{eqnarray}
From now on, all capitalized Greek letters denote two-component
spinor wave functions. In the next step, following the procedure of
Ref.~\cite{prl}, we perform the transformation to the rotating frame
to eliminate the dependence of the potential on $t-z/c$. To this
end, we make the following substitutions in Eqs.~(\ref{dirac1})
\begin{eqnarray}\label{subst}
\Psi_\pm = U{\tilde\Psi}_\pm = e^{-i(\omega t_-/2\hbar){\hat
M}_z}{\tilde\Psi}_\pm,
\end{eqnarray}
where ${\hat M}_z$ is the $z$ component of the total angular
momentum
\begin{eqnarray}\label{angmom}
{\hat M}_z = {\hat
L}_z+\frac{\hbar}{2}\sigma_z=\frac{\hbar}{i}(x\partial_y-y\partial_x)
+\frac{\hbar}{2}\sigma_z.
\end{eqnarray}
Under the transformation (\ref{subst}), the $z$ components of all
vector operators, ${\bm r}, {\bm\nabla}$, and ${\bm\sigma}$ do not
change while the $x$ and $y$ components undergo a rotation, for
example,
\begin{subequations}\label{changes}
\begin{align}
U^\dagger x U &= x \cos(\omega t_-/2) - y\sin(\omega t_-/2),\\
U^\dagger y U &= x\sin(\omega t_-/2) + y\cos(\omega t_-/2).
\end{align}
\end{subequations}
Applying these transformations to both Eqs. (\ref{dirac1a}) and
(\ref{dirac1b})  we obtain
\begin{subequations}\label{dirac2}
\begin{eqnarray}
2i\hbar\partial_+{\tilde\Psi}_+&=& {\hat h}_0\,{\tilde\Psi}_-,\label{dirac2a}\\
(2i\hbar\partial_-+\omega{\hat M}_z){\tilde\Psi}_- &=& {\hat
h}_0\,{\tilde\Psi}_+,\label{dirac2b}
\end{eqnarray}
\end{subequations}
where
\begin{eqnarray}
{\hat
h}_0=c\left[mc\sigma_z-{\bm\sigma}_\perp\!\cdot\!(i\hbar\nabla+e{\bm
A}_0)\right],
\end{eqnarray}
and ${\bm A}_0$ is the vector potential (\ref{pot}) evaluated at
$t_-=0$,
\begin{equation}\label{pot0}
{\bm A}_0=\left((B-B_0/2)y,(B+B_0/2)x,0\right).
\end{equation}
Since now there is no explicit dependence on $t$ and $z$, we may
seek solutions of Eqs.~(\ref{dirac2}) by separating these variables
\begin{eqnarray}\label{form}
{\tilde\Psi}_\pm(x,y,z,t) = e^{-i(Et-p_z z)/\hbar}\Phi_\pm(x,y),
\end{eqnarray}
where $E$ and $p_z$ are the separation constants. The equations for
${\tilde\Psi}_\pm(x,y)$ have the form
\begin{subequations}\label{dirac3}
\begin{eqnarray}
(E-p_zc)\Phi_+ &=& {\hat h}_0\Phi_-,\label{diraca}\\
(E+p_zc+\omega{\hat M}_z)\Phi_- &=& {\hat h}_0\Phi_+,\label{diracb}
\end{eqnarray}
\end{subequations}
and upon the elimination of $\Phi_+$, we obtain the following
eigenvalue equation for the transverse motion
\begin{eqnarray}\label{pauli}
\left(\!\frac{(i\hbar\nabla+e{\bm
A}_0)^2}{2m}-\frac{\hbar\omega_0}{2}\sigma_z -\frac{\Omega}{2}{\hat
M}_z\!\right)\Phi_-=E_\perp \Phi_-,
\end{eqnarray}
where the energy of the transverse motion is defined as
\begin{eqnarray}\label{transverse}
E_\perp=\frac{E^2 - m^2c^4-(p_zc)^2}{2mc^2}
\end{eqnarray}
and the effective frequency
$\Omega=\mathcal{E}\omega=(E-p_zc)\omega/mc^2$ is the same as in the
classical theory [cf. Eq.~(\ref{e})].

We shall now determine the eigenvalues and the eigenfunctions of the
Hamiltonian appearing in (\ref{pauli}). Upon the substitution of the
explicit form (\ref{pot0}) of the vector potential ${\bm A}_0$, we
obtain the following equation
\begin{eqnarray}\label{pauli1}
\Bigg[-\frac{\hbar^2}{2m}\Delta_\perp+\frac{m}{2}\left(\omega_c+\frac{\omega_0}{2}\right)^2\!\!x^2
+\frac{m}{2}\left(\omega_c-\frac{\omega_0}{2}\right)^2\!\!y^2\nonumber\\
+i\hbar\omega_c(x\partial_y+y\partial_x)  -\frac{\omega_0+\Omega}{2}{\hat L}_z\quad\quad\nonumber\\
-\hbar\left(\frac{\omega_0}{2}+\frac{\Omega}{4}\right)\sigma_z\Bigg]\Phi_-=E_\perp
\Phi_-,\quad
\end{eqnarray}
where ${\hat L}_z$ is the $z$ component of the orbital angular
momentum. This equation, by a substitution
\begin{eqnarray}\label{subst1}
\Phi_-=e^{im\omega_c xy/\hbar}\Phi,
\end{eqnarray}
is transformed to the final form
\begin{align}\label{pauli2}
\Bigg[&-\frac{\hbar^2}{2m}\Delta_\perp+\frac{m}{2}\left(\frac{\omega_0^2}{4}-\omega_c\Omega\right)x^2
+\frac{m}{2}\left(\frac{\omega_0^2}{4}+\omega_c\Omega\right)y^2\quad\quad\nonumber\\
&-\frac{\omega_0+\Omega}{2}{\hat L}_z-\hbar\left(\frac{\omega_0}{2}
+\frac{\Omega}{4}\right)\sigma_z\Bigg]\Phi = E_\perp\Phi.
\end{align}
The Hamiltonian in this equation differs only by the spin term from
the classical Hamiltonian (\ref{ham}). Since the spin part
contributes only an energy shift, we may use the classical
amplitudes of normal modes [Eq.(\ref{annih})] to construct the
annihilation operators
\begin{subequations}\label{annih2}
\begin{eqnarray}
{\hat a}_+ = \frac{1}{(\hbar N_+)^{1/2}}
\Big[t_{++}\left(\frac{\hbar}{i}\partial_y-\frac{m\Omega s_-}{1+\mu}x\right)\nonumber\;\;\\
- i\varepsilon\,t_{+-}\left(\frac{\hbar}{i}\partial_x+\frac{m\Omega s_+}{1+\mu}y\right)\Big],\;\;\\
{\hat a}_- = \frac{1}{(\hbar N_-)^{1/2}}
\Big[t_{--}\left(\frac{\hbar}{i}\partial_y+\frac{m\Omega s_+}{1+\mu}x\right)\;\;\;\nonumber\\
+ i\varepsilon\,t_{-+}\left(\frac{\hbar}{i}\partial_x-\frac{m\Omega
s_-}{1+\mu}y\right)\Big].\;\;
\end{eqnarray}
\end{subequations}
The normalization of these operators is different from their
classical counterparts by a factor of $1/\sqrt{\hbar}$ to secure the
proper commutation relations. The quantum Hamiltonian, appearing in
Eq.~(\ref{pauli2}), expressed in terms of the creation and
annihilation operators is (without the spin term)
\begin{eqnarray}\label{ham2}
{\hat H} = \hbar\Omega_+{\hat a_+}^\dagger{\hat a_+}
-{\rm sgn}(1/2+\mu)\hbar\Omega_-{\hat a_-}^\dagger{\hat a_-}\nonumber\\
+\frac{\hbar\Omega_+}{2}-{\rm sgn}(1/2+\mu)\frac{\hbar\Omega_-}{2},
\end{eqnarray}
where we have used the relations $t_{\pm+}t_{\pm-}=2\vert
1+\mu\vert\Omega_\pm$. It differs from the classical expression
(\ref{ham1}) by the usual zero-point energy terms.

Now, we shall determine a Gaussian wave function annihilated by the
operators ${\hat a}_\pm$. We may do so by substituting a Gaussian
wave function in the form
\begin{eqnarray}\label{gauss}
\psi_0=\exp(-q_xx^2/2-q_yy^2/2-iqxy)
\end{eqnarray}
into the equations
\begin{eqnarray}\label{anneq}
{\hat a}_+\psi_0=0,\;\;\;{\hat a}_-\psi_0=0
\end{eqnarray}
and find $q_x,q_y$, and $q$. By comparing the coefficients at $x$
and $y$ in (\ref{anneq}), we obtain the following set of four {\em
linear} equations for the three parameters $q_x,q_y$ and $q$
\begin{subequations}\label{overdet}
\begin{eqnarray}
q_x\vert 1+\mu\vert\,t_{+-}-q(1+\mu)\,t_{++} = m\Omega s_-t_{++}/\hbar,\\
q_y\vert 1+\mu\vert\,t_{++}+q(1+\mu)\,t_{+-} = m\Omega s_+t_{+-}/\hbar,\\
q_x\vert 1+\mu\vert\,t_{-+}+q(1+\mu)\,t_{--} = m\Omega s_+t_{--}/\hbar,\\
q_y\vert 1+\mu\vert\,t_{--}-q(1+\mu)\,t_{-+} = m\Omega
s_-t_{-+}/\hbar.
\end{eqnarray}
\end{subequations}
There exists a solution of these equations and it reads
\begin{subequations}\label{solq}
\begin{align}
q_x &= \varpi(s_++s_-)t_{++}t_{--},\\
q_y &= \varpi(s_++s_-)t_{+-}t_{-+},\\
q &= \varepsilon\varpi(s_+t_{+-}t_{--}-s_-t_{++}t_{-+}),
\end{align}
\end{subequations}
where
\begin{eqnarray}\label{lambda}
\varpi=\frac{m\Omega}{\hbar\vert
1+\mu\vert(t_{++}t_{-+}+t_{+-}t_{--})}.
\end{eqnarray}
The parameters $q_x, q_y$, and $q$ can also be expressed as the
following explicit functions of $\mu$ and $\nu$
\begin{subequations}\label{solq1}
\begin{align}
&q_x = \frac{u}{1+2\mu-4\nu}q_y,\\
&q_y = \frac{m\Omega}{4\hbar\sqrt{2\nu^2}}\nonumber\\
&\times\sqrt{(1+2\mu-4\nu)\left[(1+2\mu)(1+\mu)^2-8\nu^2-u\right]},\\
&q = \frac{m\Omega}{8\hbar\nu^2}(1+2\mu-u),
\end{align}
\end{subequations}
where $u={\rm sgn}(1/2+\mu)\sqrt{(1+2\mu)^2-16\nu^2}$.

The Gaussian wave function (\ref{gauss}) describes an analog of the
ground state of the system. This will not always be a true ground
state because for $\mu>-1/2$ the Hamiltonian is not positive
definite. Having found the state $\psi_0$ we may generate a complete
set $\psi_{mn}$ of ``excited'' states by acting on $\psi_0$ with
powers of the creation operators
\begin{eqnarray}\label{mn}
\psi_{mn}={\hat a}_+^{\dagger m}{\hat a}_+^{\dagger n}\psi_0.
\end{eqnarray}
The wave functions representing these states are Gaussians
[Eq.(\ref{gauss})] multiplied by the polynomials in the variables
$x$ and $y$ of the $(m+n)$th order. All wave functions
[Eq.(\ref{mn})] are localized around the vortex line. A complete set
of two-component wave functions can be obtained from the
one-component functions $\psi_{mn}$ by attaching two independent
two-component spinors $u_{\pm}$
\begin{eqnarray}\label{mnspin}
\Phi_{mn\pm}=\psi_{mn}u_{\pm}.
\end{eqnarray}
Choosing the spinors $u_{\pm}$ in the form $u_+=(1,0)$ and
$u_-=(0,1)$ we obtain for each choice of $m$ and $n$ two solutions
of the eigenvalue equation (\ref{pauli2}) of the same shape and
differing only in the energy eigenvalues.

Having reduced the problem to a two-dimensional harmonic oscillator,
we may use the whole arsenal of tools available in this case. We may
combine the classical trajectories with the solutions of the wave
equations and construct quantum counterparts of classical motions.
Namely, with each pair made of a classical solution of the equations
of motion described by the Hamiltonian [Eq.(\ref{ham})] and a
localized wave function [Eq. (\ref{mn})] we may associate a new
solution of the wave equation in the form of a displaced wave
function \cite{kohn,dobson,cmm} representing a localized wave packet
moving along the classical trajectory. These states correspond to
coherent states of quantum optics. We may also introduce the analogs
of optical squeezed states described by Gaussian wave functions
whose shapes are changing during the time evolution. In the next
subsection we shall use the Gaussian states to prove that the
transport of the orbits by moving vortices, clearly seen in the
classical case, carries over to the quantum case.

\subsection{Nonrelativistic quantum mechanics}

The eigenvalue equation (\ref{pauli}) does not contain the speed of
light. Therefore, one may suspect that it coincides with the energy
eigenvalue problem for a {\em nonrelativistic} charged particle in a
coordinate frame rotating with the angular frequency $\Omega/2$. We
may confirm this observation by neglecting the relativistic
corrections from the very beginning. This amounts to using the
nonrelativistic Schr\"odinger-Pauli equation instead of the Dirac
equation
\begin{eqnarray}\label{schrod}
i\hbar\partial_t\Psi=\left(\!\frac{(i\hbar\nabla+e{\bm
A}_{nr})^2}{2m}-\frac{e\hbar}{2m}{\bm\sigma}\!\cdot\!{\bm
B}_0\right)\Psi,
\end{eqnarray}
where in the spirit of the nonrelativistic approximation we have
neglected retardation in the expression (\ref{pot}) for the vector
potential
\begin{eqnarray}\label{nr}
{\bm A}_{nr}({\bm r},t)=\left(\!
\begin{array}{c}
B(y\cos\omega t-x\sin\omega t)-yB_0/2\\
B(x\cos\omega t+y\sin\omega t)+xB_0/2\\
0\end{array}\!\right)\!.
\end{eqnarray}
Note that in the Schr\"odinger-Pauli equation the magnetic moment is
coupled only to ${\bm B}_0$ because the magnetic field generated by
the vector potential (\ref{nr}) has only the constant part. After
the transformation to the rotating frame we obtain the following
eigenvalue equation
\begin{eqnarray}\label{pauli3}
\left(\!\frac{(i\hbar\nabla+e{\bm
A}_0)^2}{2m}-\frac{\hbar\omega_0}{2}\sigma_z -\frac{\omega}{2}{\hat
M}_z\!\right)\Psi=E\Psi.
\end{eqnarray}
It differs from its relativistic counterpart [Eq. (\ref{pauli})]
only by the replacement of the relativistic parameters $\Omega$ and
$E_\perp$ by their nonrelativistic limits $\omega$ and $E$. Thus,
the mathematical solutions of the eigenvalue problem obtained in the
full relativistic theory can be used, without any modifications, in
the nonrelativistic case. This does not mean that the relativistic
corrections vanish. They are all contained in a difference between
the relativistic parameters ($\Omega, E_\perp$) and the
nonrelativistic parameters ($\omega, E$) characterizing these
solutions.

We shall now apply the time dependent Schr\"odinger-Pauli equation
to the case of a moving vortex to show that the motion of a quantum
wave packet corresponds to its classical counterparts. It is
convenient to work in the gauge in which the electromagnetic field
of the wave is described by the scalar potential. This cannot be
achieved for the full electromagnetic field [Eq. (\ref{vortex})] but
it can be done in the nonrelativistic approximation, when the
retardation effects and the magnetic field of the wave are
neglected. In the case of a vortex moving on a circle, the electric
field [Eq. (\ref{vortexca})] is obtained from the following scalar
potential
\begin{widetext}
\begin{eqnarray}\label{sc_pot}
V({\bm r},t)= -eB\omega\left(\frac{x^2-y^2}{2}\cos(\omega t)+x y
\sin(\omega t) -x(x_1\cos(\omega_1 t)+y_1\sin(\omega_1
t))-y(-y_1\cos(\omega_1 t)+x_1\sin(\omega_1 t))\right).
\end{eqnarray}
Since the variable $z$ does not appear in the potential, we can
separate out the $z$-dependent part of the wave function and
consider only the following Schr\"odinger-Pauli equation describing
the motion in the $xy$ plane
\begin{eqnarray}\label{pauli4}
i\hbar\partial_t\Psi(x,y,t) =
\left(\!-\frac{\hbar^2}{2m}\Delta_\perp+\frac{m\omega_0^2}{8}(x^2+y^2)
-\frac{\hbar}{2i}\hbar\omega_0(x\partial_y-y\partial_x) +
V(x,y,t)-\frac{\hbar\omega_0}{2}\sigma_z\!\right)\Psi(x,y,t).
\end{eqnarray}
\begin{figure}[ht]
\centering {\includegraphics[width=0.7\textwidth]{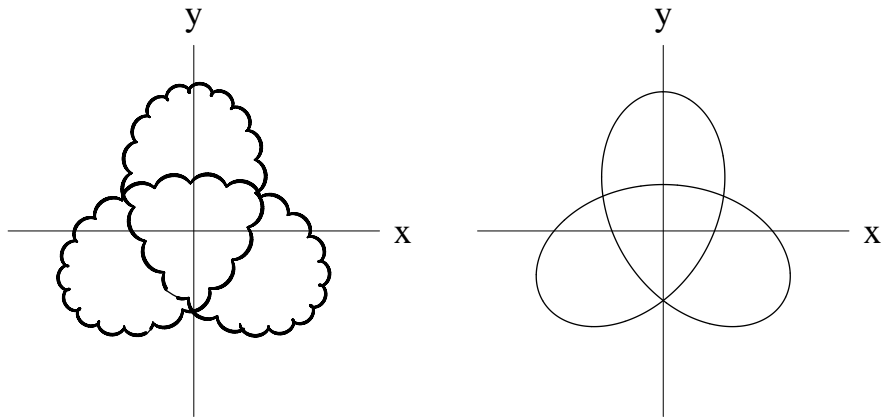}
\caption{The transport of a quantum mechanical wave packet in the
$xy$ plane by a moving vortex line. Four snapshots of the
probability distribution show the wave packet moving along the
classical trajectory (superimposed for comparison). The parameters
of the classical trajectory are the same as in Fig.~\ref{fig:8}. The
contour lines enclose the areas where the particle is found with the
probability 0.1, 0.5, 0.9, and 0.99, correspondingly. Note that due
to the influence of the vortex wave the wave packet is squeezed and
rotating.}\label{fig:10}}
\end{figure}
Upon the transformation to a frame rotating with the angular
velocity $\omega/2$, the time dependence in the quadratic part of
the Hamiltonian disappears and we obtain an equation for a
stationary driven oscillator
\begin{eqnarray}\label{pauli5}
i\hbar\partial_t\Psi(x,y,t)=
\left[\!-\frac{\hbar^2}{2m}\Delta+\frac{m\omega_c^2}{8}(x^2+y^2)-\frac{\omega_0+\omega}{2}{\hat
L}_z +  \omega_0\omega\frac{x^2-y^2}{2}-e{\bm r}\!\cdot\!{\bm
E}_1(t)-\left(\frac{\hbar\omega_0}{2}+\frac{\hbar\omega}{4}\right)\sigma_z\!\right]\Psi(x,y,t),
\end{eqnarray}
\end{widetext}
where ${\bm E}_1(t)$ is the part of the electric field in
(\ref{vortexca}) with the frequency $\omega_1$. In the absence of
the driving force ${\bm E}_1(t)$, the Gaussian solution of this
equation is given by (\ref{gauss}). In order to construct the
solution of this equation that corresponds to the classical
trajectory shown in Fig.~\ref{fig:8} we shall employ a procedure
that enables one to obtain from any given solution $\psi_0({\bm
r},t)$ of the Schr\"odinger equation with a quadratic Hamiltonian a
family of alternate solutions. These solutions are labeled by all
possible classical trajectories and are obtained by
``space-shifting'' and ``phase-shifting'' the original solution
according to the formula
\begin{eqnarray}\label{new_sol}
\psi_{(\xi,\pi)}({\bm r},t) =
e^{-iS(t)/\hbar}e^{i\pi_i(t)x^i}e^{-\xi^i(t)\partial_i}\psi_0({\bm
r},t),
\end{eqnarray}
where $\xi^i(t)$ and $\pi_i(t)$ are the classical positions and
momenta and the phase $S(t)$ is the integrated classical action.
This construction is based on a generalization of the Dobson's
harmonic potential theorem \cite{dobson}. In the Appendix we give,
for completeness, a simple proof of this theorem for the most
general quadratic Hamiltonian. We shall consider the simplest case,
when $\psi_0$ in the rotating frame is given by the Gaussian wave
function (\ref{gauss}). In the laboratory frame, the corresponding
wave function is obtained by the inverse transformation of the
variables as compared to Eq.(\ref{changes}) and it has the form
\begin{align}\label{new_sol1}
\psi_L({\bm r},t) =
e^{-q_x[x\cos(\omega t/2)+y\sin(\omega t/2)]^2/2}\nonumber\\
\times e^{-q_y[y\cos(\omega t/2)-x\sin(\omega t/2)]^2/2}\nonumber\\
\times e^{-iq[x\cos(\omega t/2)+y\sin(\omega t/2)][y\cos(\omega
t/2)-x\sin(\omega t/2)]}.
\end{align}
The new solution corresponding to the classical trajectory depicted
in Fig.~\ref{fig:8} obtained according to the prescription
(\ref{new_sol}) is
\begin{align}\label{new_sol2}
\psi_{(\xi,\pi)}({\bm r},t) = e^{-iS(t)/\hbar}e^{-q_x[x\cos(\omega t/2)+y\sin(\omega t/2)]^2/2}\nonumber\\
\times e^{-q_y[y\cos(\omega t/2)-x\sin(\omega t/2)]^2/2}\nonumber\\
\times e^{-iq[x\cos(\omega t/2)+y\sin(\omega t/2)][y\cos(\omega
t/2)-x\sin(\omega t/2)]}.
\end{align}
In Fig.~\ref{fig:10} we show the time evolution of this wave packet,
when the center of mass motion is described by the nonrelativistic
limit of the same classical trajectory that is shown in
Fig.~\ref{fig:8}. The Landau orbit is transported along a circle by
the moving vortex. This wave function describes an electron
localized with probability 0.99 inside the largest ellipse,
transported in the constant magnetic field of by an electromagnetic
wave with a moving vortex.

\section{Summary}

We have shown that classical (cyclotron) and quantum (Landau) orbits
of a charged particle in a constant magnetic field can be controlled
by electromagnetic waves with embedded vortex lines. These orbits
are pinned down in the vicinity of the vortex line and are dragged
along when the vortex line moves in the plane perpendicular to the
magnetic field. We analyzed this behavior with the use of our new
analytic solutions of the Lorentz equations of motion in the
classical case and the Dirac (or Schr\"odinger) wave equation in the
quantum case. We show that the trapping of orbits by the vortex in
the classical case has its counterpart in the form of Gaussian shape
of the wave functions localized in the vicinity of the vortex. The
effective reduction of the dynamics of our system to that of a
two-dimensional harmonic oscillator makes it possible to give a
complete solution using many tools developed in the past for
oscillators. In particular, with the use of creation operators we
have constructed a complete set of states. Coherent states, the
closest analogs of classical orbits, are also easily constructed.

\section*{Acknowledgments}
We would like to thank and Zofia Bialynicka-Birula, Dmitry Gitman,
and Tomasz Sowi\'{n}ski for useful comments and Vladislav Bagrov and
Dmitry Gitman for a copy of their book. This research was partly
supported by Grant No. 008/P03/2003, Polish Ministry of Science.

\appendix
\section{Solutions for a general quadratic Hamiltonian}

In this appendix we show how to construct the new solutions of the
Schr\"odinger equation that were used to describe the transport of
Landau orbits. Let us consider the case of the most general
quadratic Hamiltonian
\begin{eqnarray}\label{gen_ham}
H(x^i,p_i,t)=\frac{1}{2}a^{ij}(t)\pi_i\pi_j+\frac{1}{2}b_{ij}(t)\xi^i\xi^j\nonumber\\
+c_i^{\;j}(t)\xi^i\pi_j-f_i(t)\xi^i+g^i(t)\pi_i,
\end{eqnarray}
where the summation over repeated indices is understood. The
Hamiltonian equations of motion have the form
\begin{subequations}\label{ham_eq}
\begin{eqnarray}
{\dot\xi}^k(t)=a^{kj}(t)\pi_j+c_j^{\;k}(t)\xi^j(t)+g^k(t),\\
{\dot\pi}_k(t)=-b_{kj}(t)\xi^j-c_k^{\;j}(t)\pi_j(t)+f_k(t).
\end{eqnarray}
\end{subequations}
The Schr\"odinger equation for this system is ($\hbar=1$)
\begin{align}\label{schr_eq}
&i\partial_t\psi(x^i,t)=\Big[-\frac{1}{2}a^{ij}(t)\partial_i\partial_j+\frac{1}{2}b_{ij}(t)x^ix^j\nonumber\\
&-ic_i^{\;j}(t)x^i\partial_j,-f_i(t)x^i-ig^i(t)\partial_i+\frac{i}{2}c_i^{\;i}\Big]\psi(x^i,t).
\end{align}
The last term on the right hand side is added to make the quantum
Hamiltonian a Hermitian operator. For this system, the following
statement holds. For every solution $\psi_{0}(x^i,t)$ of the
Schr\"odinger equation (\ref{schr_eq}) without external forces
($f_i=0=g^i$) there corresponds a family of solutions
$\psi_{\xi\pi}(x^i,t)$ of the complete equation (\ref{schr_eq})
obtained by space-shifting and phase-shifting the original solution,
namely,
\begin{align}\label{new_sola}
\psi_{\xi\pi}(x^i,t)=e^{iS(t)}e^{i\pi_i(t)x^i}\psi_{0}(x^i-\xi^i(t),t)
\end{align}
where $(\xi^i(t),\pi_i(t))$ is any solution of the classical
equations of motion (\ref{ham_eq}). To prove this statement we
substitute into (\ref{schr_eq}) the new wave function written in the
form
\begin{align}\label{new_solb}
\psi_{\xi\pi}(x^i,t)=e^{-iS(t)}U_{\pi}U_{\xi}\psi_{0}(x^i,t),\nonumber\\
U_{\pi}=e^{i\pi_i(t)x^i},\;\;\;U_{\xi}=e^{-\xi^i(t)\partial_i}
\end{align}
and we multiply the whole equation from the left by $U^{-1}_{\pi}$,
$U^{-1}_{\xi}$, and $e^{-iS(t)}$. Next, we use the relations
\begin{align}\label{relat}
&U^{-1}_{\pi}\partial_kU_{\pi}=\partial_k + i\pi_k(t),\;\;U^{-1}_{\xi}x^kU_{\xi}=x^k + \xi^k(t),\nonumber\\
&e^{iS(t)}\partial_te^{-iS(t)}=\partial_t - i\partial_tS(t),\nonumber\\
&U^{-1}_{\pi}\partial_tU_{\pi}=\partial_t + i{\dot\pi}_i(t)x^i,\,
U^{-1}_{\xi}\partial_tU_{\xi}=\partial_t - {\dot\xi}^i(t)\partial_i
\end{align}
and we collect all the terms linear in $x^i$ and $\partial_i$ and
independent of $x^i$ and $\partial_i$. The linear terms cancel due
to the classical equations of motion (\ref{ham_eq}) and the
independent terms cancel if the phase $S(t)$ is made equal to the
following classical action integral
\begin{eqnarray}\label{phase}
S(t)=\int_0^t
dt\left(\frac{1}{2}a^{ij}\pi_i\pi_j-\frac{1}{2}b_{ij}\xi^i\xi^j+g^i\pi_i\right).
\end{eqnarray}
All remaining terms reduce to the Schr\"odinger equation satisfied
by the wave function $\psi_{0}(x^i,t)$.

\end{document}